\documentclass[%
 aip,
 amsmath,amssymb,
 reprint,%
]{revtex4-2}

\usepackage{graphicx}
\usepackage{dcolumn}
\usepackage{bm}
\usepackage[colorlinks]{hyperref}
\usepackage{subfigure}
\usepackage[final]{changes}
\usepackage[section]{placeins}
\usepackage{float}
\usepackage[mathlines]{lineno}
\usepackage{mathrsfs}

\begin{document}

\preprint{APS/123-QED}

\title{The Universe as a driven quantum system: Unbounded heating in cyclic cosmologies}%

\author{J. P. P. Vieira}
\email{ttxjp89@nottingham.ac.uk}
\affiliation{%
 Interdisciplinary Centre for Mathematical Modelling and Department of Mathematical Sciences, Loughborough University, Loughborough, Leicestershire LE11 3TU, UK
}%

\date{\today}

\begin{abstract}
The Hamiltonian of an evolving Universe is shown to be formally equivalent to that of a driven quantum system, whose driving follows from the temporal dependence of the spacetime metric. This analogy allows insights from the field of driven quantum systems to be applied to cosmological settings. In particular, it is shown that periodic cyclic cosmologies are generically prohibited (except under certain limiting constraints) due to their correspondence with periodically-driven quantum systems (which are typically expected to experience unbounded heating in the infinite future). This result highlights how future work on non-periodically-driven quantum systems is required to fully describe the dynamics of more general cyclic cosmologies (for which a qualitative picture is briefly discussed).
\end{abstract}

\maketitle


\section{Introduction}

Driven quantum systems -- quantum systems subjected to time-dependent external forces -- have been the focus of intensive study in the context of out-of-equilibrium statistical mechanics. Periodically-driven quantum systems, in particular, have attracted the most interest, owing to their tractability through the Floquet formalism \cite{PeriodicH,driven_quantum_tunneling,PGE,Eq_PeriodicDrive,MBL_periodic,MBL_periodicDrive,phase_driven}.

Here I shall begin by observing that quantum dynamics in non-static spacetimes can often be interpreted as the dynamics of a corresponding driven quantum system -- where the driving force is set by the time-dependence of the metric. In particular, this analogy holds in scenarios where quantum gravity effects are routinely considered in cosmology (such as the simplest classes of inflationary perturbations \cite{Choe_Gong_Stewart_second}). In this work, its consequences will be explored for the special case of (periodic) cyclic cosmologies.

Cyclic cosmologies have long been considered as natural devices to avoid \emph{ab initio} philosophical problems associated with the Big Bang \cite{history_world_models,history_cyclic,dynamic_cyclic}. More recently, interest in these models has been rekindled by the realisation that an ekpyrotic phase of ultra-slow contraction offers an alternative to the inflationary paradigm as a possible solution to the problems of standard Big Bang cosmology \cite{cyclic_ekpyrotic,evolution_cyclic_ekpyrotic}.

Periodic cyclic models are known to be problematic from classical considerations, as the periodic dynamics is incompatible with the inexorable increase in entropy expected from the second law of thermodynamics \cite{Tolman_book}. In this work, I identify a similar problem arising from a purely quantum effect \cite{Eq_PeriodicDrive}, which crucially differs from the classical one in that it doesn't require assumptions about thermal equilibrium and in that it applies to past as well as future evolution.

Whilst non-periodic cyclic models are not immediately subject to this result, further work on these is warranted: to establish under which conditions the physical mechanism behind this effect can be ignored; and to ascertain the fate of quantised classical periodic models (to which this work offers only an intuitive picture).

Section \ref{sec:analogy} shows how, in the appropriate limit, quantum dynamics in an evolving spacetime can be mapped into the dynamics of a driven quantum system. In Section \ref{sec:cyclic} this correspondence is explored to connect cyclic cosmologies to Floquet systems, where unbounded heating is expected -- the consequences of which, as well as conditions to avoid it, are discussed in detail in Subsections \ref{subsec:problem} and \ref{subsec:evading}, respectively. Finally, Section \ref{sec:conclusions} summarises the main conclusions of this work.

\section{The Universe as a driven quantum system \label{sec:analogy}}

Consider a Universe whose classical dynamics are dictated by the general action
\begin{equation}
S=\int dx^{4}\sqrt{-g}\left[-\frac{R}{16\pi G}+\mathcal{L}\right],
\label{eq:action}
\end{equation}
where $g$ is the determinant of the spacetime metric, $G$ is Newton's constant, $R$ is the Ricci scalar (so that the first term in brackets corresponds to the Einstein-Hilbert action), and $\mathcal{L}$ is some general Lagrangian density.

A complete theory of the quantum dynamics of this Universe would require a procedure to consistently quantise the metric terms in Eq.~(\ref{eq:action}), which is not currently known. Nevertheless, when one is interested in studying quantum effects on scales significantly smaller than the horizon length, it suffices to quantise the non-metric degrees of freedom in $\mathcal{L}$ -- as is routinely done in calculations of quantum perturbations generated during inflation \cite{STEWART_RUN,Choe_Gong_Stewart_second}.

Rather than quantise the solutions to the Euler-Lagrange equations (as is typically done), we can equivalently encode the quantum dynamics of the Universe in this approximation through a Hamiltonian with the general form
\begin{equation}
\hat{\mathscr{H}}\left(t\right)=\int d\mathrm{x}^{3}\hat{\mathcal{H}}\left(\vec{\mathbf{\mathrm{x}}},t\right),
\label{eq:H}
\end{equation}
where $\vec{\mathbf{\mathrm{x}}}$ is a three-dimensional spatial variable, $t$ is a time variable, and $\hat{\mathcal{H}}\left(\vec{\mathbf{\mathrm{x}}},t\right)$ is a quantised version of the Hamiltonian density (i.e., expressable as some generic expansion in creation and annihilation operators) which can be obtained, for example, by appropriately quantising the temporal component of the stress-energy tensor \cite{Hamiltonian_scalar}
\begin{equation}
\mathcal{H}=T_{00}=-2\frac{\partial\mathcal{L}}{\partial g^{00}}+g_{00}\mathcal{L}.
\label{eq:T00}
\end{equation}

For the special case of a Minkowski metric, $\hat{\mathscr{H}}$ is constant in the Schr{\"o}dinger picture. In more general spacetimes, however, it depends on time through temporal dependences in the metric terms in Eq.~(\ref{eq:T00}) (including eventual explicit metric dependences through covariant derivatives in $\mathcal{L}$). Therefore, the quantum dynamics of the Universe in this limit can be seen as equivalent to that of a driven quantum system whose driving is completely determined by the spacetime metric. Naturally, in general the solution for the spacetime metric may not be computable independently of the quantum dynamics. Nevertheless, it can always be formally assumed to be known \emph{a priori} (and often its dynamics can be greatly simplified by an appropriate choice of gauge).

\section{Cyclic cosmologies \label{sec:cyclic}}

\subsection{Unbounded heating in Floquet cosmology \label{subsec:heating}}

Consider now that the Universe from Eq.~(\ref{eq:action}) alternates between expanding and contracting phases. Additionally, suppose that a gauge can be chosen where this evolution is periodic. As all metric terms and their derivatives are now periodic, the Hamiltonian in Eq.~(\ref{eq:H}) is also periodic with the same period. Therefore quantum phenomena in these cyclic cosmologies can be described by the Floquet formalism for periodically-driven quantum systems \cite{PeriodicH,driven_quantum_tunneling} (by virtue of the correspondence outlined in Section \ref{sec:analogy}).

Regardless of the exact details of the driving, periodically-driven quantum systems are generically expected to evolve towards infinite-temperature steady states in the infinite future \cite{Eq_PeriodicDrive} -- in the sense that the time-averaged expectation values of an observable $\hat{O}$ are expected to converge to
\begin{equation}
\mathcal{O}\left(t\right)\longrightarrow D_{H}^{-1}\mathrm{Tr}\left(\hat{O}\right),
\label{eq:steady_state}
\end{equation}
where $D_{H}$ is the dimensionality of the available Hilbert space. In other words, quantum systems governed by periodic Hamiltonians are generically expected to converge to a state of maximum entropy.

The consequences of this result thus directly translate to an absurd realisation for these periodic cosmologies: however slowly (and it should be noted that the presence of long-range interactions is expected to make this a remarkably slow process \cite{long_range_prethermo}), the temperature in a periodic Universe is expected to increase indefinitely after a large number of periods -- which is incompatible with the original assumption that the evolution be periodic.

\subsection{A quantum heating problem for cyclic cosmologies \label{subsec:problem}}

A similar problem is known from classical considerations on the effects of the second law of thermodynamics \cite{Tolman_book}: if irreversible processes take place in the Universe (as in any realistic cosmological model), the resulting entropy increase precludes any periodic solution. Due to this unavoidable increase in entropy, the energy in a comoving fluid element increases between cycles, causing each successive oscillation to be larger (i.e., the scale factor always reaches a higher maximum value than in the previous cycle) \cite{bouncing_review}.

This problem differs in nature from the one introduced above, which comes about as the result of a purely quantum phenomenon (eigenstate mixing) and is specifically associated with driven quantum systems -- in particular not requiring any assumption about thermal equilibrium. Moreover, unlike in the classical case, the same argument can be directly applied to past evolution: meaning this heating problem is expected both in the infinite future and the infinite past (in the case of models without a beginning, or with extremely large numbers of past cycles).

Nevertheless, the classical case offers a glimpse of the actual consequences of this effect for cosmological dynamics. Extrapolating from the effect of entropy increases in that scenario, one should expect models which would classically be periodic \cite{dynamic_cyclic} to have cycles whose amplitude increases in both time directions (although more quickly in the future direction, due to the classical heating effect).


Unfortunately, there is not an obvious way to extend the Floquet formalism to more general non-periodic drivings corresponding to more general non-periodic cyclic models -- even if similar results in quasiperiodically-driven systems \cite{quasiperiodic_drive_review,bounds_quasiperiodic_drive,fine_structure_quasiperiodic_drive} suggest this may be possible in future work. However, an intuitive picture for the fate of models which would classically be periodic \cite{dynamic_cyclic} naturally emerges from the knowledge of the fate of classical models with unbounded heating. Thus, the quantum version of such models can be expected to feature a present transient (albeit extraordinarily long-lived) period of oscillatory evolution arising out of a cosmic state of nearly saturated entropy, whither the Universe returns in its eventual heat death (in a process reminiscent of the ``fluctuations'' in Eddington's discussion of the arrow of time\cite{arrow_time}).

\subsection{Evading the quantum heating problem \label{subsec:evading}}

Let us now address some ways in which the effect described in Subsection \ref{subsec:heating} may be avoided.

\subsubsection{Non-periodic cyclic models}

The argument in Subsection \ref{subsec:heating} assumes periodic evolution of the metric -- leading to a periodic Hamiltonian -- because only in that case is there a robust expectation that the resulting driven system should experience unbounded heating \cite{Eq_PeriodicDrive}. Whilst similar results have been found for quasiperiodically-driven systems\cite{quasiperiodic_drive_review,bounds_quasiperiodic_drive,fine_structure_quasiperiodic_drive} (suggesting that the class of systems which heat up in this way may be much wider), generic expectations are not known for more general drivings. As such, cyclic models which cannot be cast as periodic are natural candidates to evade this problem.

It is not difficult to write down cyclic models of this sort: combining known mechanisms by which the necessary bounces can be achieved, it is always possible to construct solutions where the scale factor oscillates in a non-periodicisable manner. In particular, cyclic models featuring ekpyrotic contraction are typically of this form -- featuring net expansion over a full cycle \cite{brane_cyclic} --  even when the evolution looks periodic to local observers \cite{new_kind_cyclic}.

Likewise, models which do not feature contraction but display cyclic Hubble parameters (such as conformal cyclic cosmology\cite{CCC} or the recent example of $\Lambda$CDM periodic cosmology\cite{periodic_LCDM}) are beyond the reach of this periodic treatment.

Whether these sorts of models are immune to this heating problem (or can be tuned to avoid it) is impossible to say without further work on the quantum front. In the absence of a more general result for driven quantum systems, the fate of each individual model may need to be considered separately (although it makes intuitive sense that this effect be negligible as long as there is sufficient net expansion over each cycle). Presumably a full analysis of this effect for Hamiltonians which classically lead to periodic evolution should reveal dynamics of the sort intuitively argued for in Subsection \ref{subsec:heating}

\subsubsection{Integrability}

The result on which this heating problem is based \cite{Eq_PeriodicDrive} does not apply if the driven Hamiltonian is integrable -- in which case the steady-state solution is given by a Periodic Gibbs Ensemble (PGE) set by the initial state \cite{PGE}. Although such solutions are compatible with periodic cosmologies, integrable Hamiltonians cannot be expected to model a realistic Universe at least during those periods of cosmic cycles when standard Big Bang cosmology applies (as required thermalisation processes are only a generically found in non-integrable systems \cite{GGE}).

\subsubsection{Spatial disorder and time crystals}

Spacial disorder in lattices has long been associated with Anderson localisation\cite{Anderson_loc}, but recently it has further been linked with the phenomenon of many-body localisation (MBL) in interacting quantum systems \cite{loc_interacting_fermions,insulator_transition_many_electrons}. Moreover, it has also been shown that the MBL behaviour of such disordered systems can (depending on the parameters of specific implementations) survive under periodic driving \cite{MBL_periodicDrive,MBL_periodic}, thus avoiding thermalisation and the fate of unbounded heating. Under certain conditions, driving these systems can result in Floquet time crystal phases\cite{Floquet_time_crystals,first_Floquet_time_crystals}, where steady states are periodic in time -- and thus natural candidates for scenarios which could be translated into feasible periodic cosmological models.

While there have been attempts to build cosmological models of (classical) time crystals\cite{cosmo_time_crystal}, it is unclear how a realistic quantum version of such models could be constructed in practice. In principle, it should be possible to come up with appropriate toy models in discrete spacetime scenarios if disorder can be introduced into its spatial structure (or alternatively if similar results can be obtained with fractal models\cite{fractal_spacetime,fractal_Floquet}). However, as with non-integrable models, these sorts of solutions (based on evading the usual thermalisation conditions) tend to preclude the usual thermalisation processes on which standard Big Bang cosmology relies; and it's not obvious that they can possibly account for such processes.

\subsubsection{Quantum gravity effects}

Naturally, any theoretical consideration about quantum effects in cosmology is potentially vulnerable to unknown features of a full theory of quantum gravity. In principle, one can always contend that the description at the basis of this work is at least as well-grounded as standard perturbation calculations in slow-roll inflation\cite{STEWART_RUN,Choe_Gong_Stewart_second} -- but it must be kept in mind that the key assumption that quantum scales of interest be small compared to cosmological ones may plausibly break down during bounces. Ultimately, the lack of strong constraints on how bouncing models may be realised means one can't completely rule out this analysis being invalidated in specific models.

\section{Conclusions \label{sec:conclusions}}

As long as quantum effects on super-horizon scales can be neglected, quantum dynamics in evolving spacetimes are equivalent to the dynamics of specific driven quantum systems. This correspondence opens the door to translating results from the field of open quantum systems into cosmologically relevant insights, and to potentially being able to use experimental realisations of such systems as analogues for cosmological scenarios.

In particular, the Floquet formalism which describes periodically-driven quantum systems thus applies to periodic cyclic cosmologies. However, by direct application of this result, the generic expectation that periodically-driven quantum systems evolve towards an infinite-temperature steady state results in the impossibility of such periodic cosmologies. 

Importantly, this problem is unrelated to the classical argument against periodic cosmologies based on the second law of thermodynamics\cite{Tolman_book}. However, since both arguments are based on heating across cycles (albeit from different sources) it is reasonable to expect that the effect of the quantum mechanism on the evolution of the scale factor should be qualitatively similar to the classical one's -- except it should be seen both in the infinite future and in the infinite past. This suggests periodic cosmologies may be salvageable as cosmological models featuring long-lived cyclic evolution arising in an otherwise steady infinite-temperature Universe.

In the end further work is needed on non-periodic driven quantum systems for a more complete picture: both to quantitatively study the evolution of models which would classically be periodic and to enable a complete description of more general cyclic models (and, in particular, to establish under which conditions these need to take into account this sort of quantum heating).

Similarly, future (periodic) cyclic models may attempt to circumvent this result by resorting to integrable dynamics, spatial disorder, or yet-unkown quantum gravity effects. However, apart from the latter, it is unclear that such ``loopholes'' in the generic expectation for Floquet systems may be exploited without affecting the system's ability to thermalise (which is necessary to recover standard cosmology).

\section{Acknowledgements}

Thanks to Achilleas Lazarides, Daniel Passos, and Luc{\'i}a F. de la Bella for useful discussions.

\bibliography{references}

\end{document}